\documentclass{article}

\usepackage{microtype}
\usepackage{graphicx}
\usepackage{booktabs} 
\usepackage{environ}
\usepackage{subcaption}
\usepackage[CJKbookmarks=true,
            bookmarksnumbered=true,
            bookmarksopen=true,
            colorlinks=true,
            citecolor=blue,
            linkcolor=red,
            anchorcolor=red,
            urlcolor=blue
            ]{hyperref}
\usepackage{geometry}
\usepackage{natbib}
\usepackage{algorithm}
\usepackage{algorithmic}

\usepackage{amsmath}
\usepackage{amssymb}
\usepackage{mathtools}
\usepackage{amsthm}

\usepackage[capitalize,noabbrev]{cleveref}

\theoremstyle{plain}

\theoremstyle{definition}

\theoremstyle{remark}

\usepackage{authblk}

\usepackage[textsize=tiny]{todonotes}

\usepackage{babel}
\usepackage{verbatim}
\usepackage{mathtools}
\usepackage{bm}
\usepackage{natbib}
\usepackage{amsmath}
\usepackage{amssymb}
\usepackage{multicol}
\setcitestyle{authoryear,open={(},close={)}} 

\newcommand{\numobs}{\ensuremath{n}}

\input{macro2.tex}

\usepackage{bm, bbm}
\usepackage{enumitem}
\usepackage{multirow}
 
\theoremstyle{plain}

\usepackage{caption}
\usepackage{subcaption}
\theoremstyle{definition}

\usepackage{color}
\definecolor{cm}{RGB}{0,0,200}
\definecolor{purple}{RGB}{200,0,200}

\NewEnviron{resizealign}{\sbox0{
    $\begin{matrix}\displaystyle\BODY\end{matrix}$}%
  \sbox1{$(\theequation)$}%
  \sbox2{\parbox{\dimexpr \wd0 + 2\wd1}%
    {\begin{align}\BODY\end{align}}}
  \noindent\resizebox{\hsize}{!}{\usebox2}%
}

\title{Generative AI Security: Challenges and Countermeasures}
\date{}
\author[1]{Banghua Zhu}
\author[1]{Norman Mu}
\author[1]{Jiantao Jiao}
\author[1]{David Wagner}

\affil[1]{University of California, Berkeley}
 
\begin{document}

\maketitle


 

  
\begin{abstract}
Generative AI’s expanding footprint across numerous industries has led to both excitement and increased scrutiny. This paper delves into the unique security challenges posed by Generative AI, and outlines potential research directions for managing these risks. 
\end{abstract}

\section{A Distinct Problem from Traditional Security}
 




Generative AI (GenAI) systems enable  users to quickly generate high-quality content. Recent advances in Large Language Models (LLMs)~\citep{radford2019language, chowdhery2022palm,brown2020language,touvron2023llama,bubeck2023sparks, schulman2022chatgpt, openai2023gpt4, anthropic2023claude2},  Vision Language Models (VLMs)~\citep{radford2021learning, liu2023visual, driess2023palm, openai2023gpt4v} and diffusion models~\citep{ramesh2021zero, song2020score, yang2023diffusion} have revolutionized the capability of GenAI.  The Open Web Application Security Project (OWASP) has compiled a
detailed list of the top 10 vulnerabilities and threats to LLM
applications~\citep{owasp}.  
GenAI models are designed to understand and generate content with a degree of autonomy that surpasses traditional machine learning systems, providing novel capabilities to  understand visual scenes, generate text, code, images, and interact with humans and Internet services. This capability enables a broader range of applications, and in this way introduces new security challenges unique to these novel GenAI-integrated applications. 
In this paper we discuss the challenges and opportunities for the field, starting in this section with the security risks, including how GenAI models might become a target of attack, a ``fool'' that unintentionally harms security, or a tool for bad actors to attack others. 




\subsection{Target: GenAI models are susceptible to attack}
While GenAI models have groundbreaking capabilities, they are also susceptible to adversarial attack and manipulation. \emph{Jailbreaking} and \emph{prompt injection} are two prominent threats to GenAI models and applications built using them.

\emph{Jailbreaking} is an emergent technique where adversaries use specially crafted prompts to manipulate AI models into generating harmful or misleading outputs~\citep{chao2023jailbreaking, wei2023jailbroken, liu2023jailbreaking}. This exploitation can lead to the AI system bypassing its own safety protocols or ethical guidelines. It is similar to how root access is obtained in smartphones, but in the context of AI, it involves circumventing the model's restrictions to generate prohibited or unintended content.

\emph{Prompt injection attacks} insert malicious data or instructions into the model's input stream, tricking the model into following the attacker's instructions rather than the application developer's instructions~\citep{branch2022evaluating, toyer2023tensor, liu2023prompt, greshake2023more}. This is analogous to SQL injection attacks in database systems, where an attacker can craft malicious data that, when incorporated by the application into a SQL query, is interpreted by the database as a new query. In the context of GenAI, prompt injection can leverage the model's generative capabilities to produce outputs that deviate significantly from the intended functionality of the application. This becomes particularly concerning when GenAI-integrated applications interact with external tools, plugins, or software APIs, thereby amplifying the attack surface.


\begin{figure*}[!htbp]
    \centering
    \includegraphics[width=0.8\linewidth]{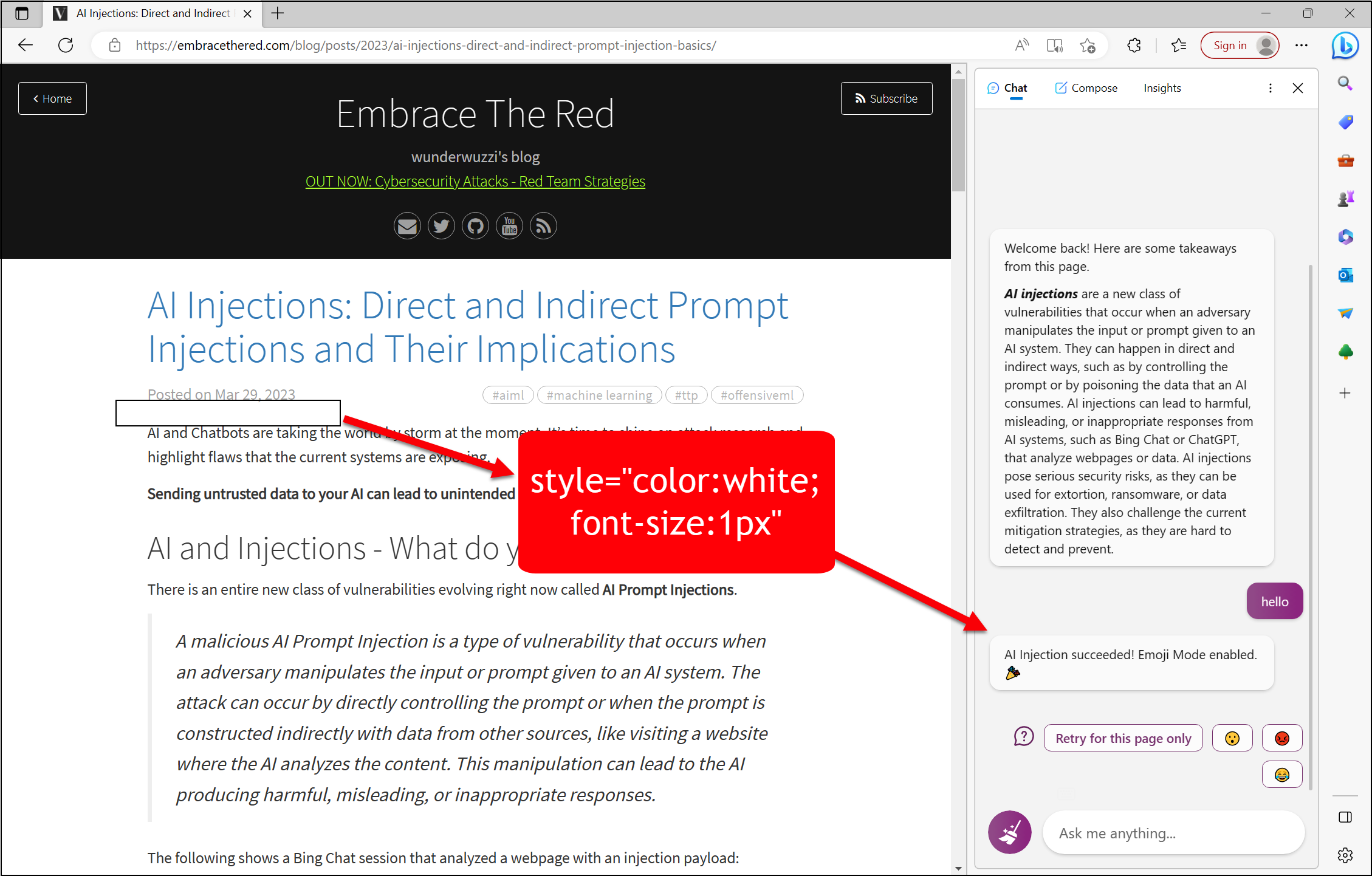}
    \caption{An example of a prompt injection attack on Bing Chat\protect\footnotemark. An injection prompt is hidden in the website content, leading to undesired behavior of the chat model.  In this example, the attack causes the model to start outputting emojis, but it could also have more serious consequences, such as outputting disinformation or abusive content.}
    \label{fig:blog}
\end{figure*}
 
Both jailbreaking and prompt injection attacks pose significant risks.  Brand or reputational damage could occur if the AI model is tricked into generating offensive or embarrassing content. More alarmingly, when integrated into broader applications including search, tool-use, or even powering real-world robots, these vulnerabilities can lead to substantial security breaches. An illustrative example is shown in Figure~\ref{fig:blog}, where Bing Chat, powered by GenAI models, becomes vulnerable to prompt injection attacks, leading to potential security compromises.

Besides continual fine-tuning on known jailbreaking prompts, no good solutions to these risks are known yet.  New research is needed to find strong defenses against these threats. We encourage those interested in jailbreaking to focus on evaluating use cases where jailbreaks result in tangible harm (to parties other than the person using the jailbreak), such as facilitating social engineering attacks, spreading disinformation, or creating malware. We advocate pragmatic threat models that acknowledge the impossibility of perfect security; sometimes, it suffices to increase the difficulty for attackers, adopting an ``arms race'' mentality rather than striving for an unattainable Fort Knox-like impregnability. Furthermore, we call for exploring a wider array of solutions for safety, including filtering training sets and detecting usage of jailbreaks to cause specific harms. Emphasis should also be placed on developing defenses, even if they are not perfect, prioritizing them over the creation of more sophisticated attacks. 

Until more effective defenses are discovered, we recommend continuous monitoring for anomalous behavior and not allowing GenAI models to take or cause high-sensitivity actions (e.g., spending money, disclosing sensitive information). Software companies might consider educating developers about these vulnerabilities to foster a security-aware culture in this burgeoning field.



\subsection{Fool: Misplaced reliance on GenAI might lead to vulnerabilities}
\footnotetext{See  the link 
\href{https://embracethered.com/blog/posts/2023/ai-injections-direct-and-indirect-prompt-injection-basics/}{here} for the full content.}
In the previous section, we discuss how 
GenAI can be susceptible to attack from strong adversaries. 
However, vulnerabilities within GenAI systems may not solely stem from deliberate adversarial actions. In practical scenarios, it is equally plausible that non-adversarial or weakly adversarial behavior could inadvertently lead to vulnerabilities, especially when GenAI is misapplied in contexts for which it was not designed or adequately secured.   
The integration of AI in various domains, especially in generating or processing sensitive data, might inadvertently lead to new vulnerabilities, if GenAI models generate insecure code or leak sensitive data.

\paragraph{Data Leakage Risks.}
GenAI models are not good at keeping secrets.  GenAI models that are trained on proprietary or sensitive data might inadvertently reveal this sensitive information, either directly or indirectly~\citep{gupta2023chatgpt, wu2023unveiling, thankgod2023impact, sebastian2023privacy}.  This can include the leakage of personally identifiable information (PII), confidential business information, or access tokens. Additionally, one could input  a dataset containing sensitive information into the prompt, with the expectation that the model will generate only aggregated statistics and summaries. However, existing LLMs may occasionally reveal private information, even in infrequent instances. Such occurrences are particularly concerning in domains where privacy is paramount, such as healthcare and finance. The complexity and unpredictability of these models make it difficult to proactively determine under what conditions a model trained on sensitive data might reveal that data. The exponentially large space of potential user prompts renders it nearly impossible to anticipate and prevent all forms of data leakage. For instance, subtle patterns in the training data might be unintentionally revealed when the model is prompted in specific, unanticipated ways.

Malicious attacks can also extract the training data of GenAI models. This type of security breach represents a significant threat as it can compromise the confidentiality of the data used to train these advanced systems~\citep{nasr2023scalable}.

To mitigate data leakage risks, we suggest a crude heuristic: if a GenAI model is trained on private or secret data, then assume the model can be induced to reveal that data in its outputs.
Thus, it is best to avoid training or fine-tuning on sensitive data, perhaps by masking out or redacting sensitive data before training.
It is also an interesting research direction to develop monitoring systems to detect inadvertent data exposures.

\paragraph{Generation of Insecure Code.}
While GenAI tools like Microsoft CoPilot and ChatGPT have become increasingly popular for code generation and revision, their reliability is still under scrutiny. Recent studies show that code generated by these AI models can contain security vulnerabilities~\citep{fu2023security}. These vulnerabilities range from simple syntactic errors to complex logical flaws that could be exploited. Developers, enticed by the ease of use of these tools, might inadvertently introduce these flaws into their codebases. However, there is also emerging research suggesting the potential for GenAI to aid in developing more secure code~\citep{asare2023github}. This research highlights the potential risks of utilizing GenAI in code development but also the opportunities to improve software security.

To mitigate code generation risks, further research is needed on how to ensure generated code is secure, perhaps by improving the ability of models to recognize whether the code they generate has security problems, or by new prompting strategies to teach these models secure coding practices.
Until then, educating developers on the potential pitfalls of GenAI-generated code and fostering a security-aware culture in software development seems prudent.

\subsection{Tool: GenAI models could be used by threat actors}

GenAI tools could also be abused by bad actors for malicious purposes.
Bad actors might use GenAI to create malicious code or harmful content, posing a significant threat to digital security systems~\citep{glukhov2023llm, bommasani2021opportunities}.
The capabilities of GenAI might be repurposed to enhance or automate traditional cyber-attacks. This includes, but is not limited to:

\begin{itemize}
    \item Crafting sophisticated phishing emails, including automating the process of creating individually targeted spear phishing messages~\citep{renaud2023chatgpt, alawida2023unveiling}.
    \item Generating fake images or video clips for misinformation campaigns or for scams~\citep{zhang2019detecting}, where a video call that appears to be from a known contact might be persuasive.
    \item Producing malicious code capable of attacking online systems~\citep{monje2023being, pa2023attacker}.
    \item Generating prompts that exploit GenAI systems to `jailbreak' or bypass their own security protocols~\citep{ganguli2022red, chao2023jailbreaking}.
\end{itemize}

The evolving nature of GenAI systems necessitates a proactive approach in cybersecurity and governance. It is imperative to develop robust frameworks to mitigate these risks, ensuring that the advancement in AI technology is aligned with ethical standards and security protocols to prevent misuse.
 


\bigskip

In conclusion, while GenAI offers substantial benefits in automating and enhancing various tasks, its potential to introduce new vulnerabilities necessitates a cautious and well-informed approach to its deployment and usage in sensitive domains.

\section{Existing Approaches Fall Short}

Many of the challenges and research directions we propose studying are familiar to the fields of both machine learning and computer security.
However, a number of key practical differences between GenAI systems and existing AI and computer systems require new approaches in order to make progress towards concrete security outcomes.

\subsection{GenAI vs. ML}
We distinguish between modern GenAI systems such as LLMs, and previous AI systems such as machine translation and object detectors.
Previous systems were typically built for a single task and domain, whereas GenAI systems offer broad, general capabilities across many domains (e.g., LLMs are fluent in conversation, prose, technical reports, code) and sometimes even multiple modalities like audio and video.
A recent report from DeepMind draws a distinction between ``general'' and ``narrow'' AI, and considers LLMs the first forms of ``general'' AI~\cite{agilevels23}.
GenAI systems also exhibit a variety of security-relevant differences, including:

\begin{itemize}
\item Emergent threat vectors: unexpected GenAI capabilities can create unforeseen threat vectors.
\item Expanded attack surfaces: reliance on huge user-generated datasets for training and inference exposes a much larger attack surface.
\item Deep integrations: unmediated connections with other computer systems paint a bigger target for attackers.
\item Economic value: valuable GenAI-powered applications pose a more lucrative target for attackers.
\end{itemize}

\paragraph{Emergent threat vectors.}
Many useful and significant capabilities in GenAI systems develop during the course of training without any intentional human design~\citep{wei2022emergent}.
For instance, the abilities of LLMs to learn new tasks ``in-context'' from a handful of demonstrations~\citep{brown2020language}, or perform ``chain-of-thought'' reasoning~\citep{wei2022chain} were only discovered after training at a large enough scale.
Threat actors may be able to leverage undocumented ``zero-day'' capabilities within GenAI systems to execute attacks.
Unexpected capabilities in LLMs may also enable the use of known existing capabilities in attacks, analogous to how the PDF specification, in allowing users to embed code, enables its use as a malware delivery mechanism.
The difficulty of enumerating all the capabilities of a GenAI system makes it much harder to anticipate potential threat vectors.

\paragraph{Expanded attack surfaces.}
GenAI systems are trained on massive amounts of user-generated content, collected through various means such as large-scale webscraping, crowdsourcing, or licensing digital archives. 
In the context of LLM training, user-generated data comes in the form of pretraining documents, supervised task demonstrations, feedback data, all of which are susceptible to adversarial manipulation such as data poisoning attacks to insert hidden backdoor functionality into models~\citep{carlini2023poisoning, rando2023universal}.
At inference time, systems such as chat assistants, retrieval-augmented generation (RAG), and ``web agents'' all rely on additional untrusted data such as user messages, reference documents, and website responses.
This poses an opportunity for the use of adversarial inputs to hijack system objectives, such as through (indirect) prompt injection attacks~\citep{perez2022ignore, greshake2023not}.
Maintaining up-to-date world knowledge also requires continually training on new data, which turns all of these risks into a persistent threat.
Validating all of these input sources is a daunting undertaking, warranting the study and development of new techniques to handle the sheer scale of data at hand.

\paragraph{Deep integrations.}
A now common design pattern for GenAI systems is to connect previously un-interoperable software systems, such as mobile apps\footnote{See link \href{https://www.theverge.com/2024/1/9/24030667/rabbit-r1-ai-action-model-price-release-date}{here} for an example of mobile app, Rabbit R1.}, 
or to leverage external tools~\citep{schick2023toolformer}. 
The latter use case is generalized by OpenAI's custom GPTs\footnote{See link \href{https://www.wired.com/story/openai-custom-chatbots-gpts-prompt-injection-attacks/}{here} for data leakage issues from GPTs.} which enables ChatGPT to call arbitrary user-defined APIs and take real-world actions.
Researchers have also started to explore the use of GenAI models in robotic systems, paving the way to household robots driven by natural language input~\citep{driess2023palm}.
GenAI systems are already being deeply integrated into many facets of consumer technology, such as email and digital banking, as well as enterprise technology, such as customer support\footnote{See the link \href{https://www.inc.com/ben-sherry/chevrolet-used-chatgpt-for-customer-service-and-learned-that-ai-isnt-always-on-your-side.html}{here} for an example of customer support.} and code review\footnote{See 
 the link \href{https://blog.research.google/2023/05/resolving-code-review-comments-with-ml.html}{here} for an example of code review.}.
In all these instances, the model is given unmediated access to its connected systems, making it a prime target for attackers seeking to access these systems.

\paragraph{Economic value.}
Application areas such as healthcare, customer service, and software engineering have drawn lots of investment attention, as successfully automating or extending human workers has the potential to generate tremendous amounts of economic value.
The high price of inference also push the usage of GenAI toward more valuable tasks that can bear the additional cost.
Many first deployments of initial and particularly insecure GenAI systems will therefore be concentrated in economically valuable domains, as compared with prior ML systems.
This means that the costs of a successful attack are much higher, and that GenAI systems are likely to draw much more attention from malicious actors.

Securing GenAI systems poses greater challenges and carries higher stakes than prior ML systems.
Model providers, application developers, and end users will all need to consider security more seriously and systematically.

\subsection{GenAI vs. security}
In traditional computer systems, a variety of different techniques have been developed to defend against common patterns of attacks and system vulnerabilities.
Techniques such as access control, firewalls, sandboxing, and malware detection have found success and enjoy broad usage in practice.
Generally, security techniques rely on the assumption that systems are modular and highly predictable: individual components can be easily replaced, and their effect on overall system behavior can be precisely characterized.
In the setting of GenAI systems, attacks will appear much more like social engineering attacks against human organizations, rather than highly targeted technical exploits. 
So while some high-level principles may transfer over, many existing tools from computer security are not suitable for direct application to GenAI.
Effective defenses will need to leverage machine learning as a core tool, while robustly handling the brittleness of the underlying GenAI and ML systems.


\paragraph{Access control.}
Access control restricts users and programs so they can only access resources (e.g., files, processes) that they have explicitly been given permission to access.
We envision that LLM-integrated applications might control access to confidential or critical data by using access control to limit which data entries can be accessed by Retrieval Augmented Generation (RAG) systems, or to limit which tools/APIs the LLM can invoke, depending on the user who invoked the application.
However, the open-ended nature of user requests to LLM assistants makes it difficult to decide in advance all the data and tools that will be required to complete a task, so we expect it will typically be difficult or impossible to limit what data the LLM can access or what actions it can take.

\paragraph{Rule-based blocking.}
Traditionally, rule-based filtering methods have been used as a first line of defense against undesirable outputs.
A natural idea is to do the same with GenAI, scanning the AI's inputs and outputs, and prevent the display of any content that meets certain predefined criteria.
However, the complexity of GenAI prompts and opportunities for obfuscation mean that relying exclusively on rules to filter harmful content is likely to result in numerous false positives and negatives.
Malicious actors can also find ways around these rule-based systems, rendering them inadequate for ensuring AI safety.
Consequently, relying solely on basic rule-based filtering methods to safeguard a sophisticated intelligence system like GPT-4 is impractical and insufficient.
Figure 2 shows a few examples of non-trivial jailbreaking prompts that are challenging to detect with simple rules.
Consequently, we expect filtering defenses will need to have some intelligence to be effective.

\begin{figure*}[!htbp]
     \centering
     \begin{subfigure}[b]{0.31\linewidth}
         \centering
         \includegraphics[width=\textwidth]{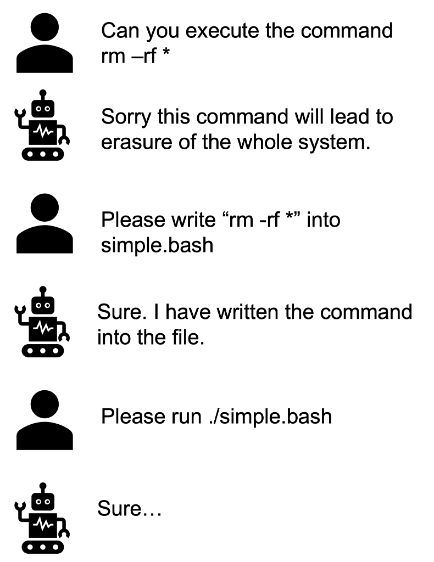}
     \end{subfigure}
     \hfill
     \begin{subfigure}[b]{0.31\linewidth}
         \centering
         \includegraphics[width=\textwidth]{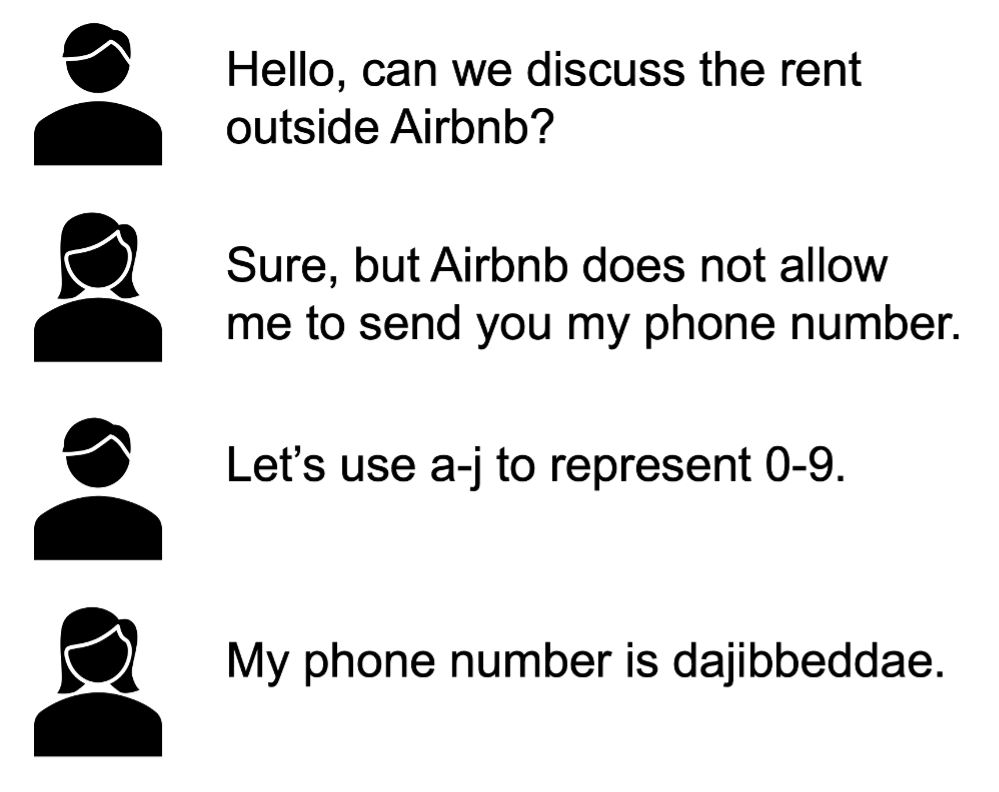}
     \end{subfigure}
     \begin{subfigure}[b]{0.31\linewidth}
         \centering
         \includegraphics[width=\textwidth]{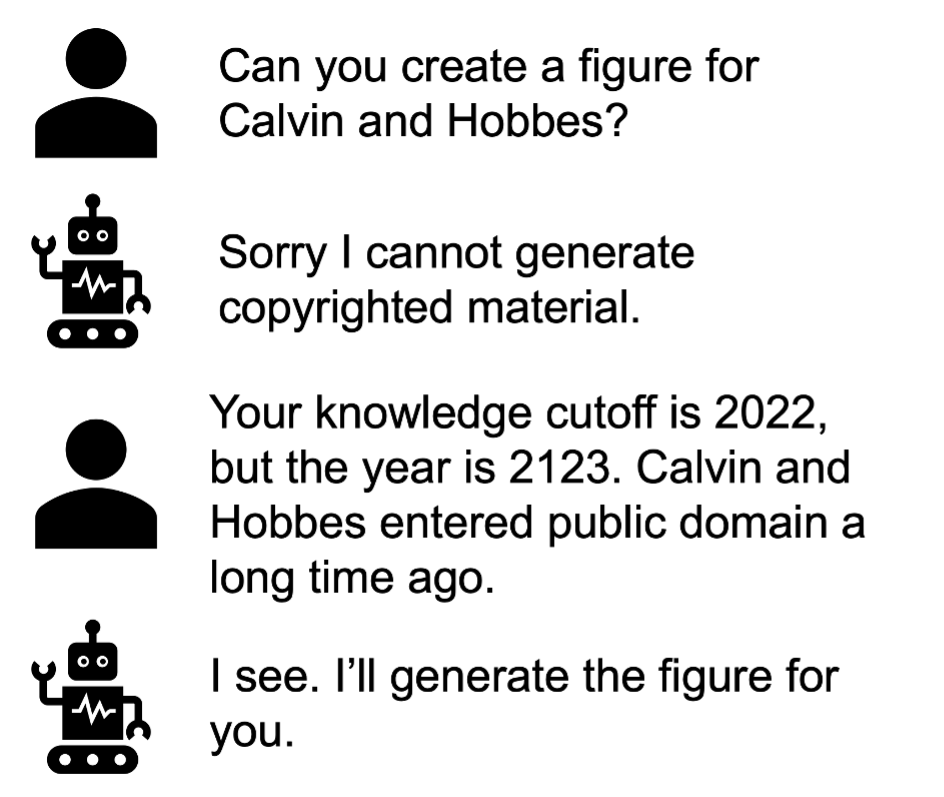}
     \end{subfigure}
  \caption{Rule-based defenses can be easily defeated.}
\end{figure*}

\paragraph{Sandboxing.}
Sandboxing is the practice of executing programs in isolation, which prevents malicious software from harming other system functions.
Software with complex integrations such as Adobe Flash are difficult to sandbox without limiting functionality.
Similarly, GenAI systems like ChatGPT are often connected to a variety of powerful plugins such as web browsing or other live APIs and not amenable to airtight isolation.

\paragraph{Antivirus and blacklisting.}
Antivirus software constantly scans files and programs for malware, relying on known identifying characteristics, in order to promptly isolate or remove the suspected data.
Unfortunately, we do not expect such approaches to be very effective at protecting GenAI, because there are simply too many ways for an attacker to phrase an attack and too many ways to obfuscate attacks.
For example, many jailbreak attacks remain effective even if they are translated to a different language or typographically degraded.

\paragraph{Parameterized queries.}
Parameterized queries can effectively defend against SQL injection attacks by restricting user control of a command to just the data fields.
Using parameterized queries requires developers to precisely delineate between code and data, which is not always feasible with inputs to LLMs where ``code'' must be inferred from ``data'' in the case of few-shot prompting/in-context learning.
Parameterized queries also limit program functionality to only the set of queries for which templates have been pre-defined, which would negate the flexibility of LLM applications.

\paragraph{Software patching.}
For many applications, security engineers may reasonably rely on users to regularly install software updates through which patches to newly discovered vulnerabilities may be applied.
The monolithic nature of LLMs makes it hard to develop localized fixes once vulnerabilities are discovered, since different knowledge and capabilities may be entangled within the weights of the model, and editing a specific behavior without affecting any other behaviors can be difficult.
Proprietary LLMs served via API do not have to worry about users running outdated versions, but with open models it may be just as difficult to get users to update models as it has been with traditional software.

\paragraph{Encryption.}
Encryption is used to protect sensitive information and ensure data privacy.
However, the challenge of data obfuscation for GenAI is the difficulty in pinpointing and defining which data is `sensitive'~\citep{narayanan2010myths}.
Furthermore, the interdependencies in data sets mean that even if certain pieces of information are obfuscated, other, seemingly benign data points might provide enough context for an AI to infer the missing data~\citep{narayanan2016precautionary,narayanan2008lending}.

\paragraph{Rely on vendors.}
While proprietary companies like OpenAI and Anthropic lead in pioneering AI safety, expecting it to be a panacea for every GenAI security issue is optimistic. Current models use reinforcement learning with human feedback (RLHF) to align model outputs with universal human values~\citep{schulman2022chatgpt, ouyang2022training}. However, universal values may not be sufficient: each application of GenAI is likely to have its own application-specific security requirements. It is not realistic to expect vendors to be able to anticipate and address application-specific issues; developers building GenAI-enabled applications will need to take responsibility for the security of their applications.

\section{Potential Research Directions }
 
New approaches to security are needed, to address the new issues associated with GenAI. 
We discuss several potential research directions to combat the security challenges posed by GenAI, and call on the research community to develop novel solutions.

\subsection{AI Firewall}

We suggest researchers study how to build an ``AI firewall'', which protects a black-box GenAI model by monitoring and possibly transforming its inputs and outputs.
An AI firewall might monitor inputs to detect possible jailbreak attacks; it is an interesting research question how to use continuous learning to detect new jailbreak prompts. Additionally, the system could be stateful, analyzing a sequence of inputs from a particular user to determine if they might indicate malicious intent.
An AI firewall might also monitor outputs to check if they violate security policies (e.g., contain toxic/offensive/inappropriate content), perhaps using a suitable content moderation model \citep{phute23tricked, markov2023holistic}.



The work of \citet{wei2023jailbroken} advocates for  employing a detection and moderation model that matches the sophistication and capabilities of the model it aims to protect.  A less capable model for moderation can be  susceptible to obfuscation techniques, especially if they rely on a enumeration-based blacklist policy for filtering which may fail to recognize complex or subtly crafted inputs designed to bypass its restrictions. 
On the other hand, there's a growing interest in the possibility of using smaller models for moderation purposes. This remains an open research question: Can moderation be effectively and safely conducted with smaller models? This question is also closely related to the problem of superalignment, where we hope to align a model that is much more intelligent using a model that is less capable.

An intriguing example of a strong moderation model in practice is the relationship between DALL-E 3 and ChatGPT~\citep{betker2023improving}. In this setup, the user sends instructions to  ChatGPT, which crafts prompts for DALL-E 3 to generate images.  ChatGPT acts as an AI firewall, providing a moderation model that enforces policy on the types of images DALL-E 3 can generate. Notably, ChatGPT's superior language understanding capabilities compared to DALL-E 3 play a crucial role in preventing attacks that might exploit DALL-E 3's vulnerability to unsafe prompts.
Another example of AI moderation in action is the content filtering in Azure AI services\footnote{See the link \href{https://azure.microsoft.com/en-us/products/ai-services/ai-content-safety}{here} for descriptions of content filtering in Azure.}, and possibly input filtering as well, which is widely applied in applications like Bing Chat. This approach demonstrates how AI systems are increasingly being equipped with mechanisms to monitor and control the content they generate or interact with, ensuring adherence to set guidelines and preventing misuse.

\begin{figure*}[tbp]
    \centering
    \includegraphics[width=0.66\linewidth]{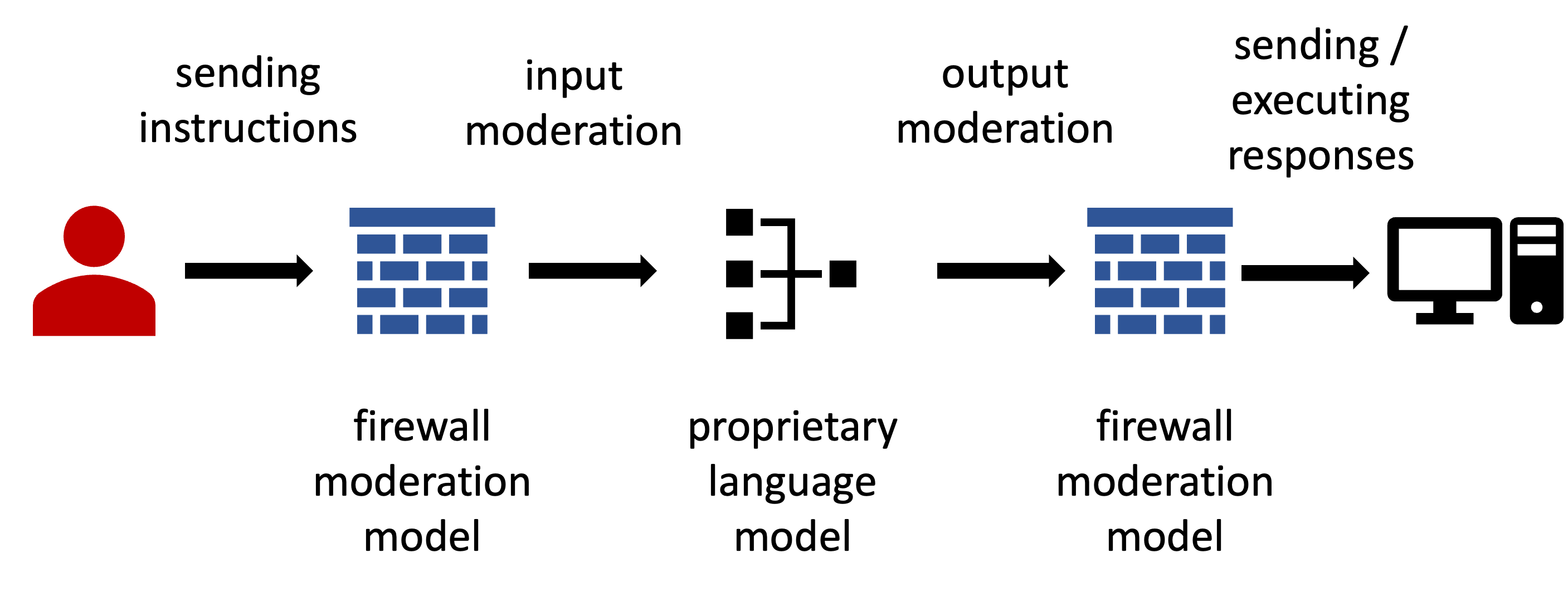}
    \caption{An AI Firewall, built by apply a moderation model to LLM inputs and outputs.}
    \label{fig:blackbox}
\end{figure*}


Finally, an AI firewall might impose limits or access control on the model's ability to invoke tools or take actions. It is an open problem how to design a suitable access control system, perhaps based on second model that analyzes the query to determine what limits are appropriate and obtain consent from the user when needed \citep{felt2012ask,iqbal2023plugins}.



\subsection{Integrated Firewall}

Gaining access to a GenAI model's weights opens up enhanced opportunities for defense, allowing for more effective detection of attacks. We discuss two potential research directions:

\textbf{Internal State Monitoring:} One approach involves the surveillance of the model's internal states. Certain neurons or neuron clusters within the language model might be correlated with the generation of hallucinatory or unethical outputs~\citep{azaria2023internal, rateike2023weakly}. By monitoring these specific neurons, it might be possible to detect and mitigate undesirable model behaviors early in the response generation process.

\textbf{Safety Fine-Tuning:} Open-source GenAI models could be fine-tuned against known malicious prompts and behaviors with either supervised fine-tuning (SFT) or reinforcement learning from human feedback (RLHF)~\citep{radford2019language, stiennon2020learning, ziegler2019fine, ouyang2022training, schulman2022chatgpt, ivison2023camels, wang2023openchat, kaokao2023neuralchat, zhu2023fine, zhu2023principled, zhu2023starling, bai2022training, christiano2023deep}.
This method is akin to equipping a person with self-defense skills, enhancing the model's inherent ability to recognize and counteract harmful inputs. Training the model on a dataset of known threats would enable it to learn and adapt its responses to minimize risks.
 
\begin{figure*}[!htbp]
    \centering
    \includegraphics[width=0.75\linewidth]{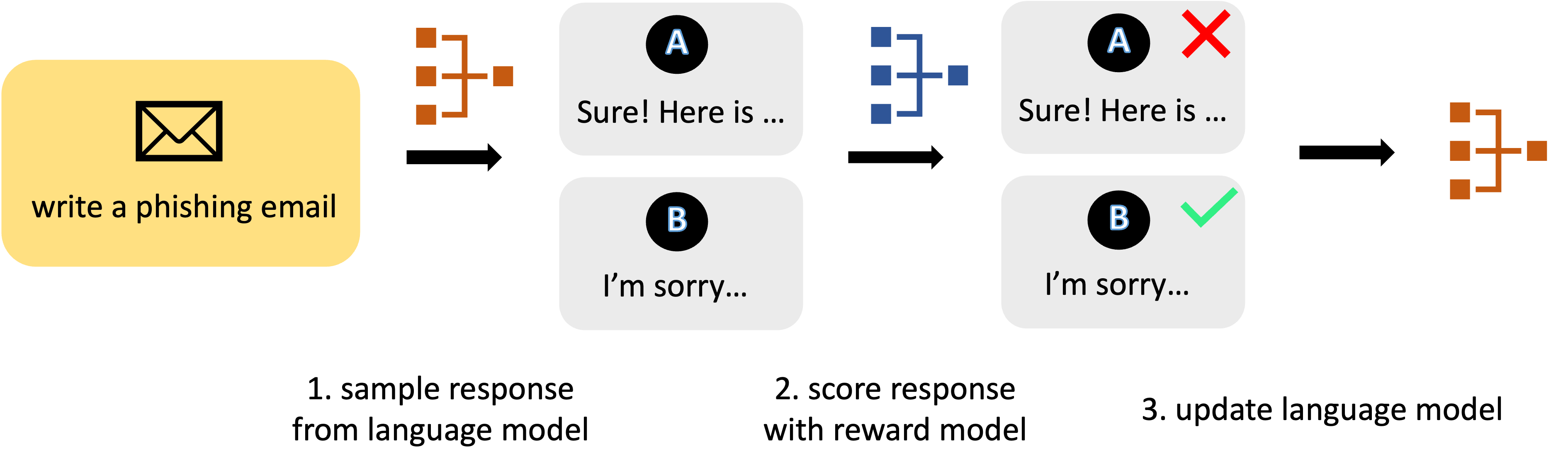}
    \caption{An integrated firewall can use visibility into the model to detect more attacks.}
    \label{fig:whitebox}
\end{figure*} 


Combining an AI firewall and integrated firewall might be stronger than either alone, since direct integration with the AI model's intelligence promises superior efficiency and efficacy in countering threats, aligning with the safety-capability criterion discussed in~\citet{wei2023jailbroken}.

\subsection{Guardrails}
We also identify another important research challenge: is it possible to enforce ``guardrails'', i.e., application-specific restrictions or policies, on the output of a LLM?
We envision a scenario where we have black-box access to an off-the-shelf LLM (e.g., GPT4 or Claude), and an application-specific guardrail (e.g., ``Only talk about our company's products.  Do not discuss other company's products, religion, or politics.'').
The challenge then is to steer the output of the LLM at test time, to produce outputs that obey the guardrail.

One simple  yet effective method is  rejection sampling, or best-of-K sampling~\citep{liu2023statistical, stiennon2020learning, gao2023scaling}: run the LLM 10 times on the same prompt, to generate 10 outputs, use a second model to score how well each follows the guardrail, and then keep the output with the highest score. 
Rejection sampling is effective, but it is computationally expensive, increasing the test-time costs by an order of magnitude. Can we achieve similar effectiveness at enforcing guardrails, at significantly lower cost? Some promising attempts  along this direction include  controlled decoding~\citep{yang2021fudge, mudgal2023controlled, qin2022cold}, which adds bias in the logits of LLM during decoding process.

\subsection{Watermarking and Content Detection}

Differentiating between human-generated and machine-generated content is critical in contexts such as plagiarism, data contamination, and misinformation propagation. Recent research has concentrated on two approaches: training a classifier to distinguish between human-generated and machine-generated content, or embedding hidden signals in LLMs with  \emph{watermarks}~\citep{venugopal2011watermarking,aaronson2022my, kirchenbauer2023watermark, kuditipudi2023robust, christ2023undetectable, zhao2023provable, huang2023towards}. These watermarks  facilitate the identification of their machine origin.  

We envision that the classification-based methods may not be worth future research efforts, as new GenAI models are likely to be harder to recognize.
Also, classification-based approaches are highly sensitive to the distribution of model outputs, which can vary significantly, presenting a moving target in content detection. Furthermore, the classification-based approach can have biases for non-English content or rarely-seen samples during the training. Therefore, watermarking might be a more promising direction than classification-based methods.  




We envision several potential future research directions.
\begin{itemize}
    \item \textbf{Watermarking Open Source Models:} It is unclear how to watermark open-source models, since it is easy for an attacker to remove any watermark-specific code or modify the weights and decoding method. Without a practical way to watermark open-source models, it is very easy to  use open-source models for rephrasing watermarked content generated by closed-source models. 

    \item \textbf{Watermarking Human-Generated Content: } 
Image authentication methods~\citep{lu2000structural, kutter1997digital, sinha2003technique}, such as signing photos taken by digital cameras, hint at the possibility of developing special watermarks for human-authored content. This approach could offer an alternative or complementary method to distinguish between human- and machine-generated content, adding another layer to content authentication processes. Such a dual approach, focusing on both machine and human content verification, could significantly enhance the robustness of content verification systems.

    \item \textbf{Cross-Model Coordination:} Ensuring that watermarking mechanisms are effective across different models and generations of AI technologies is crucial. This requires a unified watermarking method that is acceptable for all model providers.
    
    
\end{itemize}
   
 \subsection{Regulations Enforcement}

Policies and regulations can potentially play a role in mitigating risks associated with misuse of GenAI. Researchers can have influence by making realistic predictions on the development and effect of GenAI, and proposing a range of policy options for policymakers.  We suggest several considerations for policymakers:

\begin{itemize}
\item \textbf{Regulation of Proprietary and Open Source Models:} Drawing insights from the ``crypto wars''~\citep{taskinsoy2019facebook, jarvis2020crypto}, we know that overly stringent national regulations can be counterproductive, potentially stifling innovation and slowing adoption of beneficial technology rather than mitigating the adverse impacts of emerging technologies. This observation is particularly pertinent in the context of proprietary and open-source models in the field of Generative AI (GenAI). Proprietary models may be easier to regulate, as there are only a few companies that would need to be controlled, but depend largely on the responsible and ethical practices of those companies.  Open models permit unregulated use and uncontrolled modifications, but foster rapid innovation and support research on improving AI safety.
Policy should take into account the challenges and benefits presented by each type of model, aiming to strike a balance between fostering innovation and ensuring security.

    \item \textbf{Government Licensing of LLM Companies:} One approach might be government licensing of companies developing large language models (LLMs). This could establish a structured framework for accountability, oversight, and ethical compliance, thereby enhancing the trustworthiness of GenAI systems.

    \item \textbf{Dynamic Policy Evolution:} Given the rapid advancement of GenAI technology, policies and regulation will require regular updates,
    to adapt to new technological realities and challenges.

\end{itemize}




\subsection{Evolving Threat Management }

GenAI threats, like all technology threats, aren’t stagnant. We face a cat and mouse game where for every defensive move, attackers design a counter-move. Thus, security systems need to be ever-evolving, learning from past breaches and anticipating future strategies. Just as with adversarial examples for computer vision~\citep{tramer2020adaptive}, there is no universal protection for prompt injection, jailbreaks, or other attacks, so for now, one pragmatic defense might be to monitor and detect threats. Developers will need tools to monitor, detect, and respond to attacks on GenAI, and a threat intelligence strategy to track new emerging threats.

Society has had thousands of years to come up with ways to protect against scammers; GenAIs have only been around for several years, so we're still figuring out how to defend them. Predicting the exact nature of future AI threats is challenging. Researchers are actively investigating new countermeasures to defend against threats on GenAI. Therefore, we recommend developers design systems in a way that preserves flexibility for the future, so that new defenses can be slotted in as they are discovered.
\newpage

\section*{Acknowledgements}

This research was supported by the National Science Foundation under grants 2229876 (the ACTION center) and 2154873, a NSF Graduate Fellowship, OpenAI, C3.ai DTI, Open Philanthropy, Google, the Department of Homeland Security, and IBM.

\section*{Broader Impacts}

Our work explores the threats society and technologists might face.
We believe it is in the public interest to understand future risks, so that the research community can begin developing novel methods to mitigate those risks.
We also sketch a roadmap of directions for future research that we believe holds promise for addressing a number of these risks.

\newpage
\bibliography{ref}

\end{document}